# AN ADAPTIVE WATERMARKING PROCESS IN HADAMARD TRANSFORM

PARVATHAVARTHINI S[1]   SHANTHAKUMARI R[2]

[1]Department of Information Technology, Kongu Engineering College, Perundurai, Erode.
[2]Department of Information Technology, Kongu Engineering College, Perundurai, Erode.

*ABSTRACT*

*An adaptive visible/invisible watermarking scheme is done to prevent the privacy and preserving copyright protection of digital data using Hadamard transform based on the scaling factor of the image. The value of scaling factor depends on the control parameter. The scaling factor is calculated to embedded the watermark. Depend upon the control parameter the visible and invisible watermarking is determined. The proposed Hadamard transform domain method is more robust again image/signal processing attacks. Furthermore, it also shows that the proposed method confirm the efficiency through various performance analysis and experimental results.*

## I. INTRODUCTION

With the popularity of Internet communications and the development of multimedia technology, digital media can be easily duplicated, distributed and tampered. Copyright protection and content authentication are the problem faced by the digital media. Digital watermarking technique provides an approach to deal with these problems. Watermarking is the process of inserting data into a multimedia element such as an image, audio, or video file. The embedded data can be identified or extracted from the multimedia for identifying the copyright owner. An effective watermarking scheme should meet certain requirements including transparency, robustness, security, un-ambiguity, adequate information capacity, low computational complexity and etc.

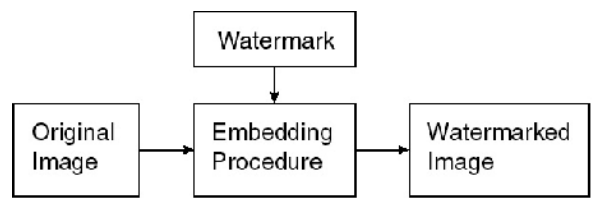

Figure 1. General watermarking method

Watermarking can be of two types Visible Watermarking and Invisible Watermarking. A visible watermark is a visible semi-transparent text or image overlaid on the original image. An invisible watermark is which cannot be perceived with human's eyes when the image is embedded. Digital Watermarking mainly involves two phases Watermark embedding, and Watermark extraction. Digital watermark can be classified into two classes depending on the domain of watermark embedding spatial domain and transform domain. In the transform domain the properties of the underlying domain can be exploited. The transform domain scheme is typically more robustness towards to noise, common image processing, and compression when compared with the spatial transform scheme. The two most commonly used methods are Discrete cosine transform (DCT), Discrete wavelet transform (DWT), Discrete fourier transform, Hadamard transform etc., Hadamard transform is consider as the one of the best method because transformation matrix is always represented in + 1 and -1. The processing time is reduced in this method since it uses only addition and subtraction operation. It has a good energy compaction property and its transformation matrix could be generated using fast algorithms and hence it is called Fast Hadamard Transform (FHT).

## 2. PROPOSED WORK

In proposed system, adaptive visible/invisible watermarking scheme for digital images is done through hadamard transform by calculating scaling factor. The value of the scaling factor is based on control parameter. It can be adjusted to make the watermark scheme as either visible or invisible. This work consist of three modules,

- Hadamard Transformation
- Watermark Embedding
- Watermark Extraction

### 2.1 Hadamard transformation

The Hadamard transform (also known as the Walsh–Hadamard transform) is orthogonal transformation that decomposes a signal into a set of orthogonal, rectangular waveforms called Walsh -Hadamard transform. It is a non-sinusoidal signal. The 2D-Hadamard transform is mainly used in image processing and image compression applications.

$$H_{2N} = \frac{1}{\sqrt{2N}} \begin{bmatrix} H_N & H_N \\ H_N & -H_N \end{bmatrix} \qquad (1)$$

The advantages of Hadamard transform include
  (i) Its elements are real
  (ii) Its rows and columns are orthogonal to each other

Let [I] represents the original image and [$\bar{I}$] is the transformed image then the 2D – Hadamard transform of [I] is given by

$$[\bar{I}] = \frac{H_N [I] H_N}{N} \qquad (2)$$

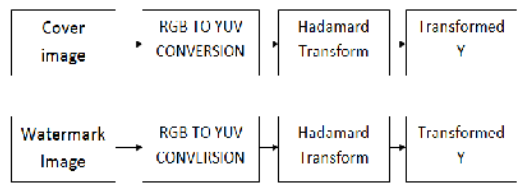

Figure 2. Process of Hadamard Transform

## 2.2. Watermark embedding

After the Hadamard transform, watermark is embedded based on the scaling factor of the cover image. The cover image to be protected or to be watermarked is transformed into frequency components using Hadamard transform technique.

$$\mu(I) = \frac{1}{P \times Q} \sum_{a=1}^{P} \sum_{b=1}^{Q} I(a,b) \qquad (3)$$

$$\alpha_1 = \frac{1}{(1+e^{-\mu(I)})} \qquad (4)$$

The   can be acquired by multiplying it with inverse multiples of 10 as given in  . The watermarking scheme is determined by the controlling parameter m and it can take the values from 0 onwards controls the strength of scaling factor.

$$= \alpha_1 \times \frac{1}{10^m} \qquad (5)$$

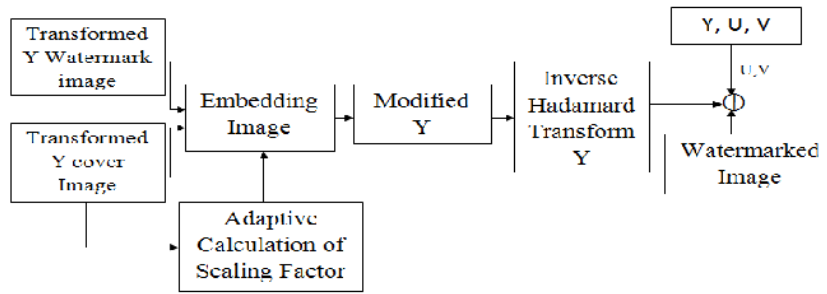

Figure 3. Watermark embedding process

For m=0 the watermark dominates the cover image and destroy the underlying image quality heavily. For m=1 the visibility of the watermark is good without destroying the underlying content. For m=2 the watermark become invisible without degrading the quality of underlying digital image.

## 2.3. Watermark Extraction

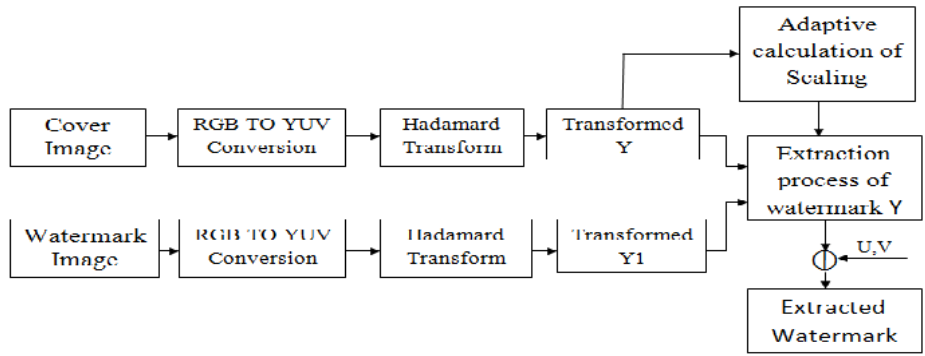

Figure 4. Watermark extraction Process

The given original cover image and watermarked image are taken as input and converted into YUV color space. Hadamard transformation technique is applied on the luminance channel of cover image and watermarked image. Hadamard transformed cover image is used to calculate the scaling factor and watermark is extracted.

Table 1 Calculated scaling factor for sample cover images

| Sample cover images | Mean value of cover images | Calculated scaling parameter for selected value of controlling parameter | | |
|---|---|---|---|---|
| | | $m=0$ | $m=1$ | $m=2$ |
| F-16 | 0.71341 | 0.5005 | 0.05000 | 0.0050 |
| Lena | 0.50287 | 0.5006 | 0.05000 | 0.0050 |

## 3. PERFORMANCE ANALYSIS

The performance for an adaptive visible/invisible watermarking scheme have been evaluated and compared on the basis of two measures: Universal Image Quality Index (UIQI), and Structural Similarity (SSIM) they are computed as follows

### 3.1. Universal Image Quality Index

It is another metric for measuring the quality of watermarked image. It is designed to model image distortion using three factors namely loss of correlation, luminance distortion and contrast distortion. For any two images x and y, the UIQI is given is

$$Q = \frac{\sigma_{xy}}{\sigma_x \sigma_y} \cdot \frac{2\bar{x}\bar{y}}{(\bar{x})^2 + (\bar{y})^2} \cdot \frac{2\sigma_x \sigma_y}{\sigma_x^2 + \sigma_y^2}. \tag{5}$$

$$[I] \quad\quad [II] \quad\quad [III]$$

Where,

[I] - correlation between x and y and is in the range between [-1,1]
[II] - closeness of luminance is between x and y range is [0,1]
[III] - similarity of contrast between x and y and is in the range of [0,1]

The universal image quality index Q lies in the range [-1, 1] where 1 indicate that x and y are similar and -1 represents the dissimilarity.

### 3.2. Structural Similarity index

It is regularly used metric for measuring the similarity between two images. It is defined as

$$\text{SSIM}(x,y) = \frac{(2\mu_x \mu_y + c_1)(2\sigma_{xy} + c_2)}{(\mu_x^2 + \mu_y^2 + c_1)(\sigma_x^2 + \sigma_y^2 + c_2)} \tag{6}$$

Where,

$\mu_x$ - the average of x

International Journal of Advanced Information Technology (IJAIT) Vol. 4, No. 2, April 2014

$\mu_y$ - the average of y
$\sigma_x^2$ - the variance of x
$\sigma_y^2$ - the variance of y
$\sigma_{xy}$ - the covariance of x and y
$c_1 = k_1 L^2$ - constant to avoid instability where $\mu_x^2 + \mu_y^2$ is close to zero; $k_1 = 0.01, L = 225$
$c_2 = k_2 L^2$ - constant to avoid instability where $\sigma_x^2 + \sigma_y^2$ is close to zero; $k_1 = 0.01, L = 225$

## 4. RESULTS AND DISCUSSION

In this part experimental and performance analysis results are given with detail. The proposed work is implemented on Intel dual core with 1GB RAM configured using MATLAB R2012a.The most commonly used test Images are taken for experimentation.

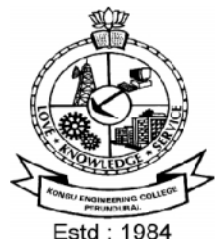

Figure 5. Watermark image

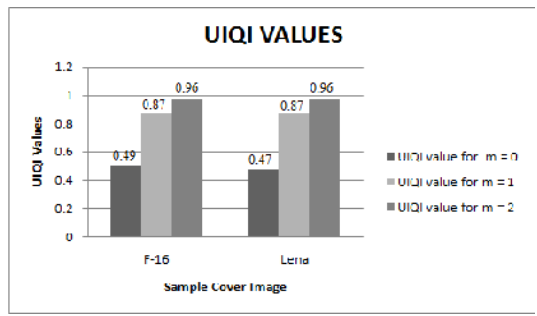

Figure 6. UIQI based evaluation for visible and invisible watermarked images.

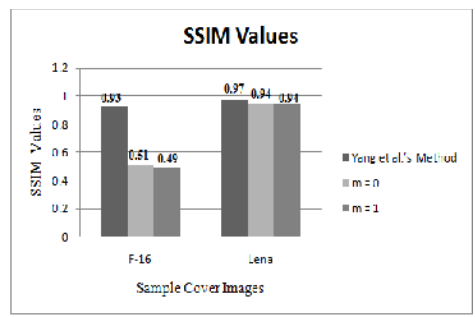

Figure 7. Obtained SSIM value for different sample values and compared with yang et al's method.

International Journal of Advanced Information Technology (IJAIT) Vol. 4, No. 2, April 2014

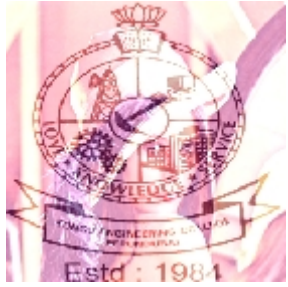    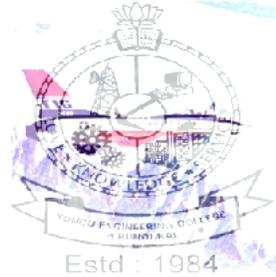

(a) Watermarked Lena  = 0.5005           (b) Watermarked F-16  = 0.5006

Figure 8. Watermark sample images for controlling parameter for m=0

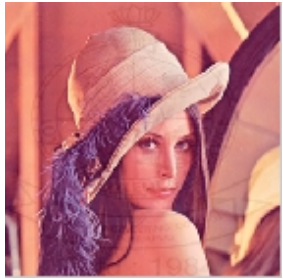    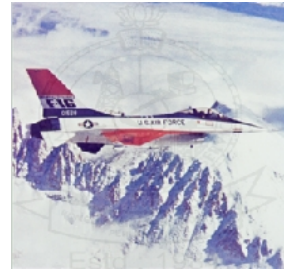

(a) Watermarked Lena  = 0.05000          (b) Watermarked F-16  = 0.0500

Figure 9. Watermark sample images for controlling parameter for m=1

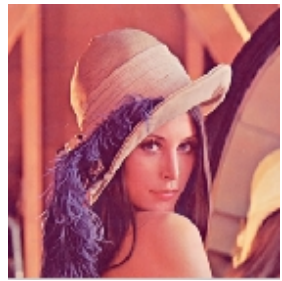    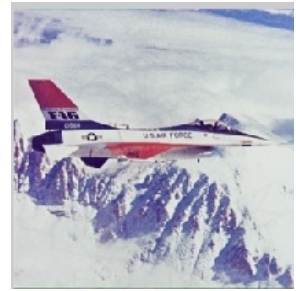

(a) Watermarked Lena  = 0.005            (b) Watermarked Lena  = 0.00500

Figure 10. Watermark sample images for controlling parameter for m=2

Table 2 UIQI based evaluation for visible and invisible watermarked image

| Sample Cover Images | UIQI value for m = 0 | UIQI value for m = 1 | UIQI value for m = 2 |
|---|---|---|---|
| F-16 | 0.49463 | 0.87020 | 0.96492 |
| Lena | 0.47377 | 0.87246 | 0.96774 |

Table 3 Comparison of obtained SSIM values with results yang et al.'s method

| Sample Cover Images | Yang et al.'s Method | SSIM Value for m = 0 | SSIM Value For m = 1 |
|---|---|---|---|
| F-16 | 0.93 | 0.51154 | 0.49031 |
| Lena | 0.97 | 0.94586 | 0.94374 |

## 5. CONCLUSION

An adaptive watermarking method based on embedding and extracting a digital watermark into/from an image is done using a hadamard transform. Hadamard transform is mainly used for its robustness against image processing attacks. In this method an adaptive procedure for calculating scaling factor or scaling strength is used in hadamard transform. The scaling factor value is controlled by a control parameter. As per the users' requirements the inserted watermark may be either visible/invisible based on the control parameter. The results are compared with the existing system. The proposed scheme confirms the efficiency by the experimental result and performance analysis.